\documentclass[aps, prd, preprint, amsfonts, floatfix]{revtex4}
\usepackage{amsmath,amssymb,array,mathrsfs,amsfonts}
\usepackage{epsfig}
\usepackage{euscript}

\usepackage{graphicx}
\usepackage{amsmath,amsfonts}

\begin{document}
\newcommand{\be}{\begin{eqnarray}}
\newcommand{\ee}{\end{eqnarray}}
\newcommand\del{\partial}
\newcommand\nn{\nonumber}
\newcommand{\Tr}{{\rm Tr}}
\newcommand{\T}{{\rm T}}
\newcommand{\noi}{{\noindent}}

\title{Full simulation of chiral Random Matrix Theory at non-zero chemical potential by Complex Langevin}
\author{A. Mollgaard}
\affiliation{Niels Bohr Institute, University of Copenhagen, Blegdamsvej 17, 2100 Copenhagen {\O}, Denmark}
\author{K. Splittorff}
\affiliation{Discovery Centre, Niels Bohr Institute, University of Copenhagen, Blegdamsvej 17, 2100 Copenhagen {\O}, Denmark}

\date   {\today}
\begin{abstract}
It is demonstrated that the complex Langevin method can simulate chiral 
random matrix theory at non-zero chemical potential. The successful 
match with the 
analytic prediction for  the chiral condensate is established through 
a shift of matrix integration variables and choosing a polar 
representation for the new matrix elements before complexification. 
Furthermore, we test the proposal
to work with a Langevin-time dependent quark mass and find that 
it allows us to control the fluctuations of the phase of the fermion 
determinant throughout the Langevin trajectory. 
\end{abstract}
\setlength{\parskip}{10pt}

\maketitle

\section{Introduction}

First principles non-perturbative simulations of full QCD have 
been limited to the region with small ratio 
of the quark chemical potential over temperature or heavy quark masses  
because of the fermion sign problem, for reviews see 
\cite{Aarts:2013bla,deForcrand:2010ys,Splittorff:2006vj}. 
Recently, however, complex Langevin
simulations of full QCD at non-zero chemical potential have been 
presented \cite{Sexty:2013ica,Sexty:2014dxa,Aarts:2014bwa}. While 
these initially are carried out in specific parameter domains 
the method holds the possibility to provide first principles 
simulations for any value of the chemical potential, even with  
low temperature and light quark masses.  

The introduction of chiral random matrix theory 
\cite{EdwardJac,JacIsmeal,JacThreefold} at non-zero chemical 
potential \cite{misha,O} has lead to a number of analytic 
insights into the non-perturbative dynamics of dense strongly 
interacting matter and the effect of the sign problem 
\cite{misha,O,phase-diag,SV-phase,OSV}. In 
\cite{MS} chiral random matrix theory was used to emphasize 
the potential problem which complex Langevin faces in simulations 
of QCD at low temperature and light quark masses.

In this paper we demonstrate that the complex Langevin approach can 
solve the sign problem in chiral random matrix theory. This is 
relevant for QCD at low temperature and light quark masses  
since chiral random matrix theory 
\cite{EdwardJac,JacIsmeal,JacThreefold,misha,O} and 
full QCD share a number of properties: 
The integral formulation of the partition function in both cases 
includes the determinant of a Dirac operator and the flavor symmetries 
and explicit breaking hereof are identical. 
QCD and chiral random matrix theory, therefore, have the 
same low energy theory in the microscopic limit, namely chiral 
perturbation theory at leading order in the $\epsilon$-domain 
\cite{EdwardJac,Basile:2007ki,VW,Akemann:2007rf}. Moreover, the 
anti-Hermiticity of the 
Dirac operators are in both cases broken by the chemical potential. 
This is particularly relevant for the present study, since it implies 
that the average of the phase factor of the fermion determinant in 
QCD and chiral random matrix theory have the same analytic form of the 
exponential suppression in the limit of large volume/size of the matrix
 \cite{SV-phase}.
In other words the sign problem in QCD for light quarks at low temperature 
and in chiral random matrix theory are equally severe.

In the physical domain where chiral random matrix theory and QCD 
share the same low energy limit, both partition functions are 
independent of the chemical potential $\mu$. This is natural as
the partition function is dominated by pions which have zero 
quark charge. The measures in the 
partition functions are however strongly $\mu$-dependent. The 
numerical problem of realizing the $\mu$-independence in both cases 
becomes particular challenging once the chemical potential exceeds 
half of the pion mass ($\mu>m_\pi/2$), see eg.~\cite{SV-phase}.

In this paper we show that complex Langevin provides a method to 
numerically simulate chiral random matrix theory, even in the domain 
of $\mu>m_\pi/2$. The success compared to the first study \cite{MS} 
is established through 
a shift of the matrix variables in the integrant as well as using a polar 
representation for the new matrix elements.

The advantage of simulating chiral random matrix theory compared to 
full QCD is that we have 
exact analytical solutions to test the numerics against. Tests of 
this kind are imperative since complex Langevin is not guarantied 
to provide the correct solutions, see for example 
\cite{Ambjorn:1986fz,aartsXY}. One potential problem particularly 
relevant for QCD is that the fermion determinant renders the action 
non-holomorfic \cite{MS}, existing 
proofs of correct convergence \cite{Aarts1,Aarts2,Aarts3} therefore 
do not apply directly.

As demonstrated in \cite{MS,Greensite} complex Langevin may lead 
to wrong results if the argument of the logarithm (the fermion determinant 
in QCD and in chiral random matrix theory) frequently circles 
the origin. 
In \cite{DiracCL} a practical proposal was given in order to 
circumvent this problem: By initially decreasing the quark mass
with the Langevin time it was suggested that it might be possible 
to reach the desired value of the quark mass without frequent 
circulations of the origin 
by the  fermion determinant. Here we test this proposal and show that 
it is indeed the case within chiral random matrix theory.

The results are presented as follows: In the next section the chiral 
random matrix theory is defined and the relevant analytic results 
are stated. Then in section \ref{sec:CL} the parametrization of the 
matrix elements is chosen and we explicitly compute the 
Langevin drift. Section \ref{sec:simulations} presents the numerical 
results obtained and the proposal to work with an initially Langevin 
time dependent quark mass is tested. We draw conclusions 
and provide an outlook in section \ref{sec:conc}.

\section{chiral Random Matrix Theory}
\label{sec:chRMT}

The chiral random matrix theory we will simulate with complex Langevin 
has the partition function \cite{O,canonical}
\be
Z_N^{N_f}(m) & = & \int  d\Phi_1 d\Phi_2 \; {\det}^{N_f}\left(D_\mu+m\right)e^{-2N\Tr[\Phi_1^\dagger\Phi_1+\Phi_2^\dagger\Phi_2]},
\label{ZRMT}
\ee
where the random matrix analogue of the Dirac operator is
\be
D_\mu+m=\left(\begin{array}{cc}
m & e^\mu \Phi_1 - e^{-\mu} \Phi_2^\dagger\\
-e^{-\mu} \Phi_1^\dagger+e^\mu \Phi_2 & m
\end{array}\right).\label{D+m}
\ee
The integration variables $\Phi_1$ and $\Phi_2^\dagger$ are complex $N\times (N+\nu)$ 
matrices, $m$ and $\mu$ are the quark mass and chemical potential parameters 
and $N_f$ is the number of quark fields which have been integrated out. The 
integer $\nu$ is the topological index, i.e.~the number of exact zero 
eigenvalues of $D_\mu$.
In the microscopic limit where $mN$ and $\mu^2N$ are fixed as $N\to\infty$ 
this random matrix partition function is equivalent to that of chiral 
perturbation theory in the $\epsilon$-domain 
\cite{O,Basile:2007ki,Akemann:2007rf}. This limit also allow us to 
identify the relation between the random matrix parameters $N$, $m$ 
and $\mu$ and the physical four volume, quark mass and chemical potential, 
 \cite{O,SV-fact,AOSV}
\be
2mN \leftrightarrow m\Sigma V \ \ \ \ {\rm and} \ \ \ 2\mu^2N \leftrightarrow \mu^2F_\pi^2V,
\ee    
where $\Sigma$ is the chiral condensate and $F_\pi$ is the pion decay constant.
In the quenched and the phase-quenched theories a phase transition takes 
place at $\mu=m_\pi/2$. Using the Gell-Mann - Oakes - Renner relation we 
can rewrite this as $\mu^2F_\pi^2V=m\Sigma V/2$, which in the chiral random 
matrix variables translates to $2\mu^2=m$.

For the numerical test of complex Langevin below we will naturally work 
with finite $N$. It is therefore of great practical value that the partition 
function (\ref{ZRMT}) can be computed analytically for all values of $N_f$ 
and $N$ \cite{O,canonical} 
\be
Z_N^{N_f}(m) & = & \frac{1}{(2 m)^{1/2 N_f (N_f - 1)}} \det\left[
   \left(\frac{d}{dm}\right)^a L_{N + b}^{(\nu)}(- n m^2)\right]_{a=0,\ldots,N_f - 1;\; b=0,\ldots,N_f-1}, 
\ee
where $L_k^{(\nu)}(x)$ is the generalized Laguerre polynomial. 
From this compact expression for the partition function we obtain 
the mass dependent chiral condensate
\be
\Sigma_N^{N_f}(m) =\frac{1}{N_f}\frac{1}{N}\frac{1}{Z_N^{N_f}(m)}\frac{d}{dm}Z_N^{N_f}(m). 
\label{Sigma}
\ee
Note that, as discussed in the introduction, the partition function is 
independent of the chemical potential even though the weight in the 
integral representation (\ref{ZRMT}) is heavily $\mu$ dependent.

\subsection{Complex Langevin dynamics}
\label{sec:CL}

When the action is complex it is natural to generalize real Langevin 
dynamics \cite{Damgaard:1987rr} by complexifying the fields and 
define a complexified Langevin dynamics \cite{Parisi,Klauder:1983nn}
through the gradient of the action.

In order to set up the complex Langevin dynamics for the chiral random 
matrix theory we first express the partition function (\ref{ZRMT}) 
\be
Z_N^{N_f}(m) & = & \int  d\Phi_1 d\Phi_2 \; \exp\left(-S\right),
\label{ZS}
\ee
in terms of the action
\be
S=2N{\rm Tr}[\Phi_1^\dagger\Phi_1+\Phi_2^\dagger\Phi_2]-N_f{\rm Tr}\left[\log\left(m^2-X Y\right)\right],
\ee
where
\be
X &\equiv& e^\mu\Phi_1 - e^{-\mu}\Phi_2^\dagger \\
Y &\equiv& -e^{-\mu}\Phi_1^\dagger+e^\mu\Phi_2 . \nn
\ee
Note the appearance of the logarithm of the fermion determinant in the action.

Next we choose to parameterize the elements of the 
complex $N\times (N+\nu)$ matrices $\Phi_1$ and $\Phi_2$ as
\be
(\Phi_1)_{ij}=r_{1,ij}e^{i\theta_{1,ij}} \quad (\Phi_2)_{ji}=r_{2,ji}e^{i\theta_{2,ji}},
\label{para}
\ee
where $i=1,\ldots,N$ and $j=1,\ldots,N+\nu$. 
In the complex Langevin dynamics the $4N(N+\nu)$ real variables $r_{1,ij}$, 
$\theta_{1,ij}$, $r_{2,ij}$, $\theta_{2,ij}$ will be complexified.  
The motivation for the choice of parametrization (\ref{para}) is that the 
$\mu$-independence of the partition function can be achieved if the Langevin 
process in effect shifts the integration contour of the $\theta_{1,ij}$ and 
$\theta_{2,ij}$ variables by $i\mu$ into the complex plane while the radial 
variables $r_{1,ij}$ and $r_{2,ij}$ are attracted to the real axis. 

In the parameterization (\ref{para}) the Gaussian term is simply 
\be
\Tr\left[\Phi_1^\dagger\Phi_1+\Phi_2^\dagger\Phi_2\right] & = & \sum_{ij} \ r_{1,ij}^2+r_{2,ji}^2,
\ee
and the action is thus
\be
S & = &-\sum_{ij} \log(r_{1,ij})+\log(r_{2,ji}) -\log\det(m^2-XY)
 +2N\sum_{ij} r_{1,ij}^2 + r_{2,ji}^2,
\ee
with
\be
X_{ij} & = & e^{\mu+i\theta_{1,ij}}r_{1,ij}-e^{-\mu-i\theta_{2,ji}}r_{2,ji}
\label{XYij} \\
Y_{ij} & = & -e^{-\mu-i\theta_{1,ji}}r_{1,ji}+e^{\mu+i\theta_{2,ij}}r_{2,ij}.\nn
\ee
The $\log(r_{1,ij})$ and $\log(r_{2,ji})$ terms are from the Jacobian for the 
change to polar variables.  

The Langevin dynamics is given by the equations 
\be
r_{1,mn}^{(t+1)}&=&r_{1,mn}^{(t)}-\frac{\partial S}{\partial r_{1,mn}}dt+\sqrt{dt}\;\eta(t) \\
r_{2,mn}^{(t+1)}&=&r_{2,mn}^{(t)}-\frac{\partial S}{\partial r_{2,mn}}dt+\sqrt{dt}\;\eta(t) \nn \\
\theta_{1,mn}^{(t+1)}&=&\theta_{1,mn}^{(t)}-\frac{\partial S}{\partial \theta_{1,mn}}dt+\sqrt{dt}\;\eta(t) \nn \\
\theta_{2,mn}^{(t+1)}&=&\theta_{2,mn}^{(t)}-\frac{\partial S}{\partial \theta_{2,mn}}dt+\sqrt{dt}\;\eta(t) , \nn 
\ee
where the derivatives are to be evaluated at Langevin-time $t$ and 
$\eta$ is a real Gaussian white noise 
\be
\big\langle \eta(t)\eta(t')\big\rangle = 2 \delta(t-t').
\ee

The next step is to compute the detailed form of the drift terms which enters 
the Langevin equations. With the notation
\be
P^{-1}\equiv(m^2-XY)^{-1}
\ee
we obtain
\be\label{driftth1}
-\frac{\partial S}{\partial \theta_{1,mn}} & = & -N_f\left[\left(P^{-1}\right)_{ki}\partial_{\theta_{1,mn}}\left(X_{ij}Y_{jk}\right)\right]\nonumber \\
 & = & -N_f\left[P^{-1}_{ki}\left(ie^{\mu+i\theta_{1,mn}}r_{1,mn}\delta_{mi}\delta_{nj}Y_{jk}+iX_{ij}\delta_{nj}\delta_{mk}e^{-\mu-i\theta_{1,mn}}r_{1,mn}\right)\right]\nonumber \\
 & = & -iN_f\left(e^{\mu+i\theta_{1,mn}}r_{1,mn}P^{-1}_{km}Y_{nk}+e^{-\mu-i\theta_{1,mn}}r_{1,mn}P^{-1}_{mi}X_{in}\right)\nonumber \\
 & = & -iN_f\left[e^{\mu+i\theta_{1,mn}}r_{1,mn}\left(\left(YP^{-1}\right)^\T\right)_{mn}+e^{-\mu-i\theta_{1,mn}}r_{1,mn}\left(P^{-1}X\right)_{mn}\right] ,
\ee
while for the radial variable we have
\be\label{driftr1}
-\frac{\partial S}{\partial r_{1,mn}} 
& = & -4Nr_{1,mn}+1/r_{1,mn}-N_f\left[\left(P^{-1}\right)_{ki}\partial_{r_{1,mn}}\left(X_{ij}Y_{jk}\right)\right] \\
 & = & -4Nr_{1,mn}+1/r_{1,mn}-N_f\left[P^{-1}_{ki}\left(e^{\mu+i\theta_{1,mn}}\delta_{mi}\delta_{nj}Y_{jk}-X_{ij}\delta_{nj}\delta_{mk}e^{-\mu-i\theta_{1,mn}}\right)\right]\nonumber \\
 & = & -4Nr_{1,mn}+1/r_{1,mn}-N_f\left(e^{\mu+i\theta_{1,mn}}P^{-1}_{km}Y_{nk}-e^{-\mu-i\theta_{1,mn}}P^{-1}_{mi}X_{in}\right)\nonumber \\
 & = & -4Nr_{1,mn}+1/r_{1,mn}-N_f\left[e^{\mu+i\theta_{1,mn}}\left(\left(YP^{-1}\right)^\T\right)_{mn}-e^{-\mu-i\theta_{1,mn}}\left(P^{-1}X\right)_{mn}\right] \nn .
\ee
Similarly for the angular and radial variables from $\Phi_2$ we have
\be\label{driftth2}
-\frac{\partial S}{\partial \theta_{2,mn}} & = & -N_f\left[\left(P^{-1}\right)_{ki}\partial_{\theta_{2,mn}}\left(X_{ij}Y_{jk}\right)\right] \\
 & = & -N_{f}\left[P^{-1}_{ki}\left(ie^{-\mu-i\theta_{2,mn}}r_{2,mn}\delta_{ni}\delta_{mj}Y_{jk}+iX_{ij}\delta_{mj}\delta_{nk}e^{\mu+i\theta_{2,mn}}r_{2,mn}\right)\right]\nonumber \\
 & = & -iN_f\left(e^{-\mu-i\theta_{2,mn}}r_{2,mn}P^{-1}_{kn}Y_{mk}+e^{\mu+i\theta_{2,mn}}r_{2,mn}P^{-1}_{ni}X_{im}\right)\nonumber \\
 & = & -iN_f\left[e^{-\mu-i\theta_{2,mn}}r_{2,mn}\left(YP^{-1}\right)_{mn}+e^{\mu+i\theta_{2,mn}}r_{2,mn}\left(\left(P^{-1}X\right)^\T\right)_{mn}\right] , \nn
\ee
and
\be\label{driftr2}
-\frac{\partial S}{\partial r_{2,mn}} 
 & = & -4Nr_{2,mn}+1/r_{2,mn}-N_f\left[\left(P^{-1}\right)_{ki}\partial_{r_{2,mn}}\left(X_{ij}Y_{jk}\right)\right] \\
 & = & -4Nr_{2,mn}+1/r_{2,mn}-N_f\left[P^{-1}_{ki}\left(-e^{-\mu-i\theta_{2,mn}}\delta_{ni}\delta_{mj}Y_{jk}+X_{ij}\delta_{mj}\delta_{nk}e^{\mu+i\theta_{2,mn}}\right)\right]\nonumber \\
 & = & -4Nr_{2,mn}+1/r_{2,mn}-N_f\left(-e^{-\mu-i\theta_{2,mn}}P^{-1}_{kn}Y_{mk}+e^{\mu+i\theta_{2,mn}}P^{-1}_{ni}X_{im}\right)\nonumber \\
 & = & -4Nr_{2,mn}+1/r_{2,mn}-N_f\left[-e^{-\mu-i\theta_{2,mn}}\left(YP^{-1}\right)_{mn}+e^{\mu+i\theta_{2,mn}}\left(\left(P^{-1}X\right)^\T\right)_{mn}\right]. \nn 
\ee
Note that we have ignored the cut of the logarithm and simply used the
standard form for the derivative of the log when the argument is real 
and positive. This has potential consequences for the Langevin process 
\cite{MS} in particular if the argument of the log frequently circles 
the origin of the complex plane. We will return to this point in  
section \ref{subsec:moft} below.

The Langevin dynamics presented above differs in two ways from that 
used in \cite{MS}. First, the realization (\ref{ZRMT}) of the chiral random 
matrix theory partition function is related to the partition function 
used in \cite{MS} (see Eq.~(4.1) therein) by a change of the matrices 
which enters as integration variables (see also the appendix of 
\cite{canonical}). Second, the parametrization of the matrix elements 
(\ref{para}) used here is different from that used in \cite{MS}. Both 
changes are nessesary for the success of the simulations presented below.

\label{sec:simulations}
\begin{center}
\begin{figure}[t!]
\includegraphics[width=10.5cm,angle=0]{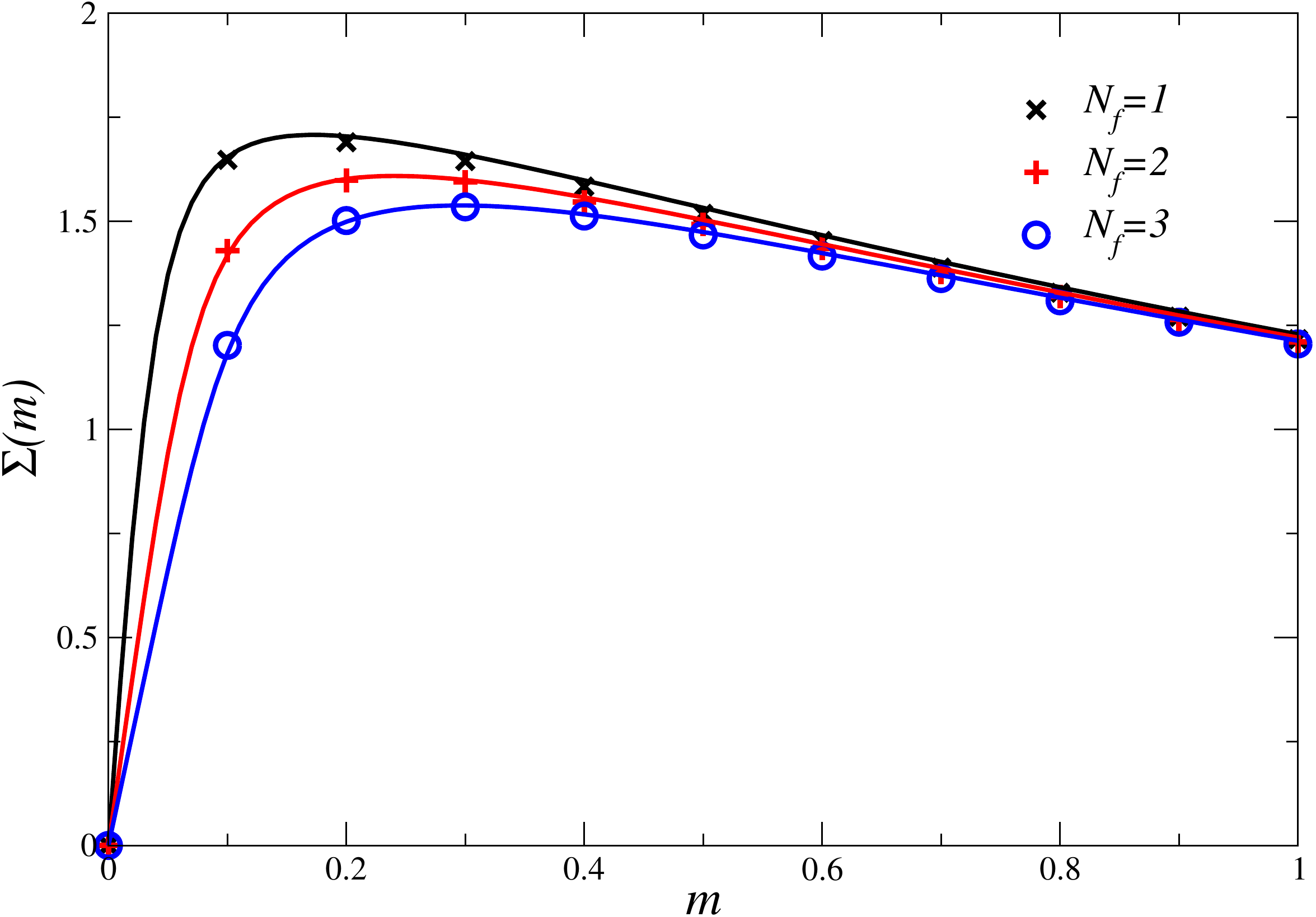}
\caption{\label{fig:SigmaNf} The chiral condensate as a function of the quark mass for one, two and three dynamical flavors. 
The full lines are the exact analytic predictions and the points are the results of the complex Langevin dynamics with adaptive step-size. The parameters chosen for the plot are $N=20$, $\mu=1$, $\nu=0$ and $2T=2000$.}
\end{figure}
\end{center}

\section{Simulations}

In order to test the complex Langevin algorithm presented above 
we have run a series of numerical simulations. The central observable 
of interest is the chiral condensate for which the analytic prediction 
is given in Eq.~(\ref{Sigma}). This observable is 
not only relevant for the non-perturbative physics it is also 
highly sensitive to the sign problem, since the phase-quenched 
chiral condensate takes a drastically different form in the region 
of $m<2\mu^2$, see eg.~\cite{MS} for plots hereof.

To compute the chiral condensate we start the Langevin process 
in a random configuration from the original (not complexified)  
quenched ensemble. The Langevin process is then run for a period, $2T$, in 
time-steps of $dt$ and on the latter half of the trajectory we compute   
\be
\Sigma(m) = \frac{1}{N_f}\frac{1}{N}\frac{1}{T}\sum_{t=T+dt}^{2T} 
{\rm Re}\,\Tr\,\frac{1}{D_\mu^{(t)}+m} \; dt\;,
\ee
where $D_\mu^{(t)}+m$ is the Dirac operator (\ref{D+m}) evaluated for 
the complexified fields generated through (\ref{driftth1})-(\ref{driftr2})
at Langevin time $t$.
The first half, $T$, of the period allows the Langevin process to react to the initial condition. 
\begin{center}
\begin{figure}[t!]
\includegraphics[width=10.5cm,angle=0]{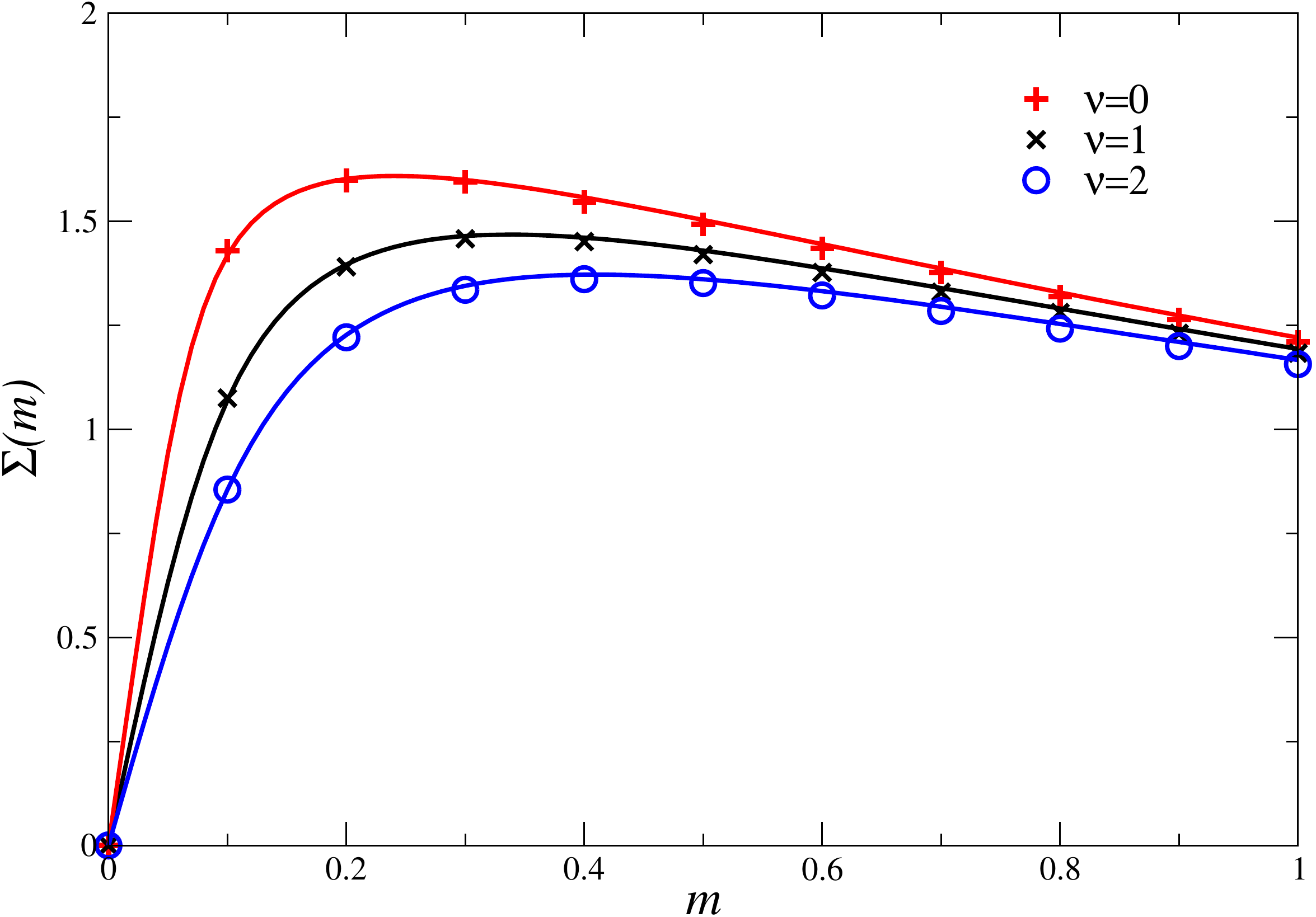}
\caption{\label{fig:Sigmanu} The chiral condensate as a function of the quark mass for $\nu=0,1,2$ with $N_f=2$, $N=20$, $\mu=1$ and $\nu=0$. The data points are obtained with complex Langevin using adaptive step-size and the full lines are the predictions (\ref{Sigma}).}
\end{figure}
\end{center}
\begin{center}
\begin{figure}[t!]
\includegraphics[width=10.5cm,angle=0]{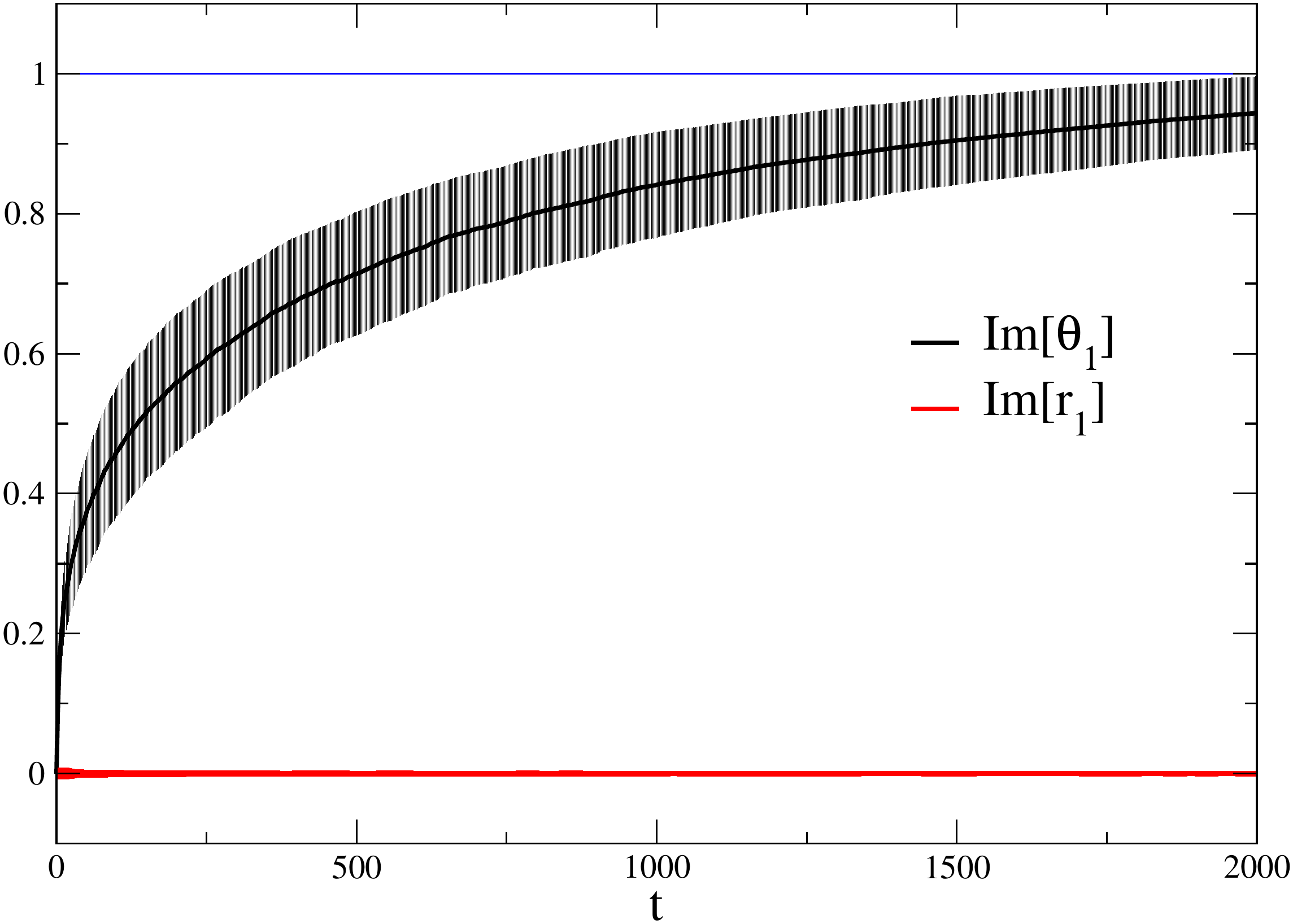}
\caption{\label{fig:th_t} The average of the imaginary part of the angular variable $\theta_1$ and the average of the imaginary part of the radial variable $r_1$. As a function of the Langevin time $t$ the angular variable flows towards $\mu=1$ marked by the thin vertical line while the radial variable remains close to the real axis. The error bars are given by the square root of the variance. The parameters in the simulation are $\nu=0$, $N_f=2$, $N=20$, $\mu=1$, $m=1$ and $\nu=0$.}
\end{figure}
\end{center}
\begin{center}
\begin{figure}[t!]
\includegraphics[width=10.5cm,angle=0]{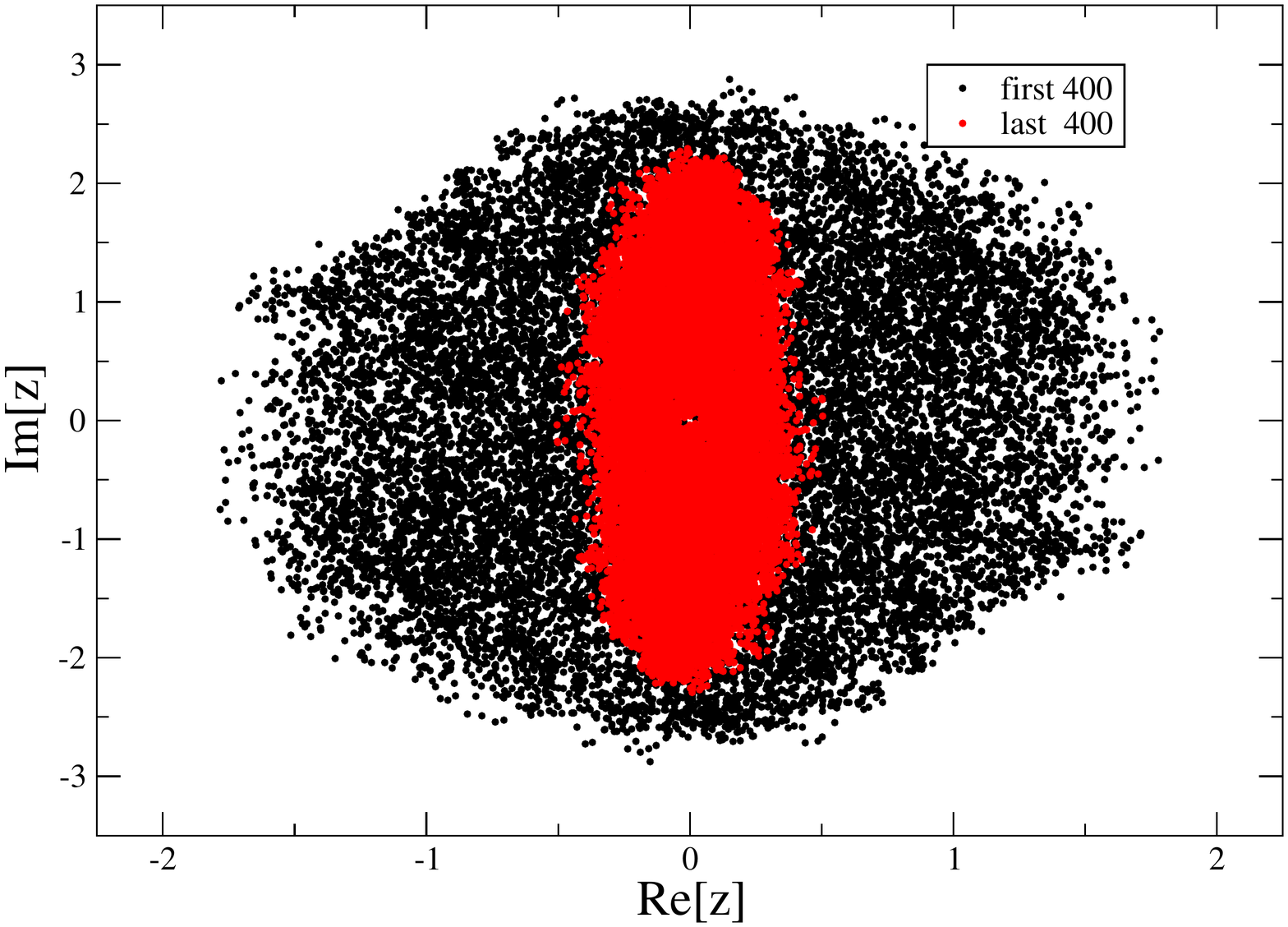}
\caption{\label{fig:Dirac} The eigenvalues of the Dirac operator in the 
complex plane {\bf black} for the first 400 time steps {\bf red} for the 
final 400 time steps. Parameters used are $N=20$, $N_f=2$, $\mu=1$, $m=1$, $\nu=0$ and $60000$ adaptive steps. Note that the quark mass is initially well inside the eigenvalue distribution.} 
\end{figure}
\end{center}
\vspace{-38mm}
With increasing size of the matrices  
we have found it convenient to implement adaptive step-size \cite{Ambjorn:1986fz,adaptive}, 
since the 
$1/r$-terms in the drift can lead to large excursions unless $dt$ is 
sufficiently small. Results for the chiral 
condensate for $N=20$, $\mu=1$, $2T=2000$, $\nu=0$ and adaptive step-size 
are shown in figure 
\ref{fig:SigmaNf}.
Displayed are the numerical results for $N_f=1$, $2$ and $3$ as well as   
the analytic predictions. We observe that the Langevin 
process is able to reproduce the expected mass dependence 
throughout the range of values for $m$ with all three values of $N_f$. 
Note that $m<2\mu^2$ throughout the range displayed. The convergence 
is equally good for larger values of $m$. 

Next we have tested the Langevin dynamics for non-zero topological index 
$\nu$. In figure \ref{fig:Sigmanu} the numerical results for $\nu=0$, $1$ 
and $2$ with adaptive step-size, $N_f=2$, $N=20$, $2T=2000$ are plotted against 
the analytic curves. The numerical data again follow the expected curves
and demonstrates that the topological zero modes causes no obstacle for 
the complex Langevin algorithm in chiral random matrix theory.

In order to gain insights in the dynamics of these successful simulations 
we have monitored the values of the variables throughout the Langevin 
process: The angular variables $\theta_{1,ij}$ and $\theta_{2,ij}$  
are effectively shifted by $i\mu$ into the complex plane while the 
$r_{1,ij}$ and $r_{2,ij}$ are attracted toward the real axis. This 
cancels the $\mu$-dependence of the chiral condensate, as was indeed  
the motivation for the choice of parametrization (\ref{para}). An example 
of the flow of the variable is shown in figure \ref{fig:th_t}. The band is 
the average of the imaginary part of the elements in $\theta_1$ with the 
errors given by the squareroot of the variance. 

\begin{center}
\begin{figure}[t!]
\includegraphics[width=10cm,angle=-90]{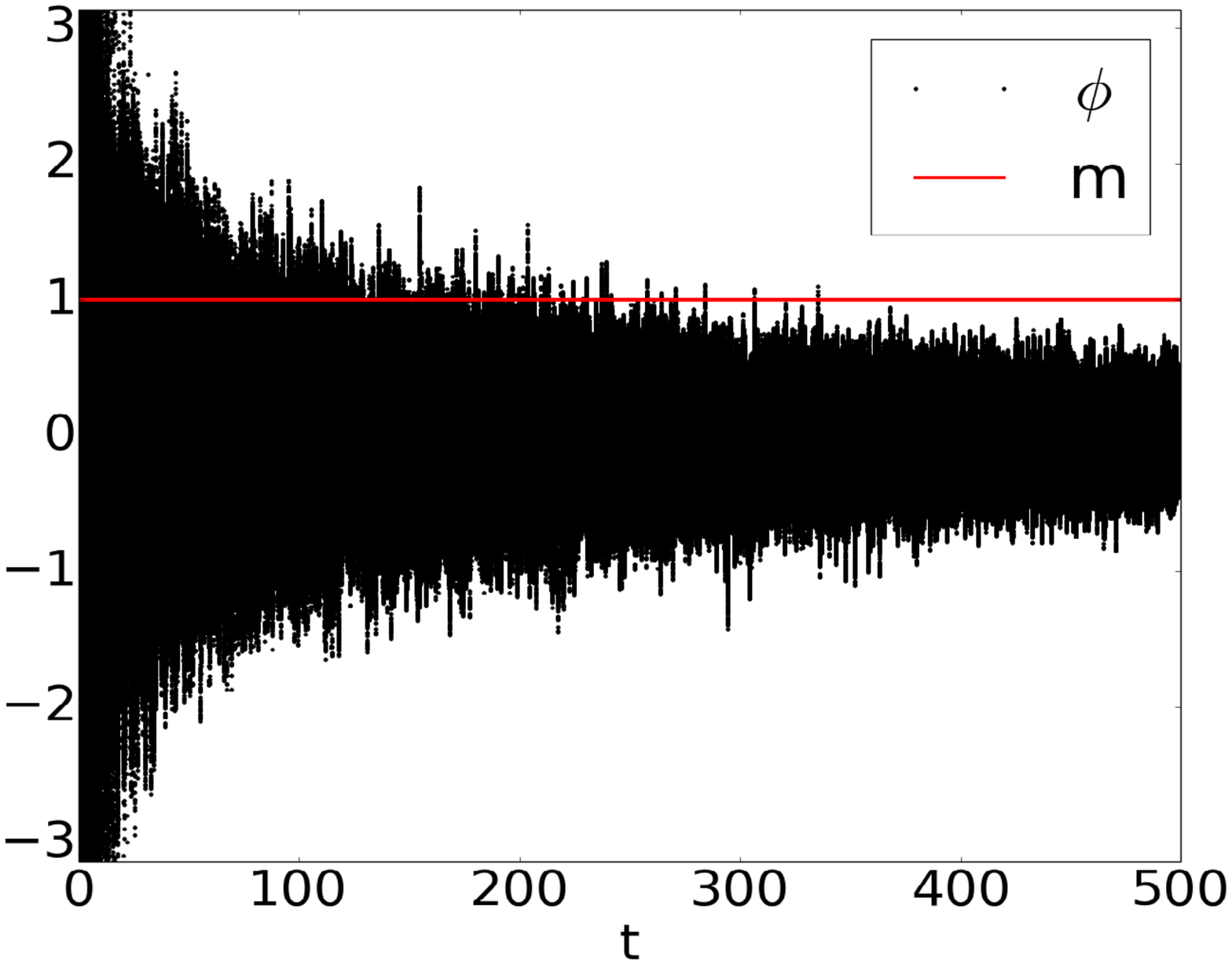}
\caption{\label{fig:m-fixed} The argument, $\phi$, of the phase of the fermion determinant 
as a function of Langevin time, for $m=1$, $\mu=1$, $N=20$, $N_f=2$ and 
$\nu=0$. Initially the fermion determinant frequently circles the origin but 
with the Langevin time the Dirac eigenvalues flow inside the quark mass and 
the fluctuations of the fermion determinant are damped. The fixed value of the 
quark mass throughout the run is indicated bu the horizontal red line.}
\end{figure}
\end{center}

The flow of the variables manifest themselves also in the distribution 
of the Dirac eigenvalues. In figure \ref{fig:Dirac} the eigenvalues of 
the Dirac operator on the initial 400 configurations are plotted in black 
along with the eigenvalues of the final 400 configurations out of 
60000 adaptive steps.
The value of the quark mass, $m=1$, is well within the initial eigevalue 
distribution, however, with Langevin time the Dirac eigenvalues moove inside 
the quark mass. As this happens the fluctuations of the phase of the 
fermion determinant are damped, see figure \ref{fig:m-fixed}.

\begin{center}
\begin{figure}[t!]
\includegraphics[width=10cm,angle=-90]{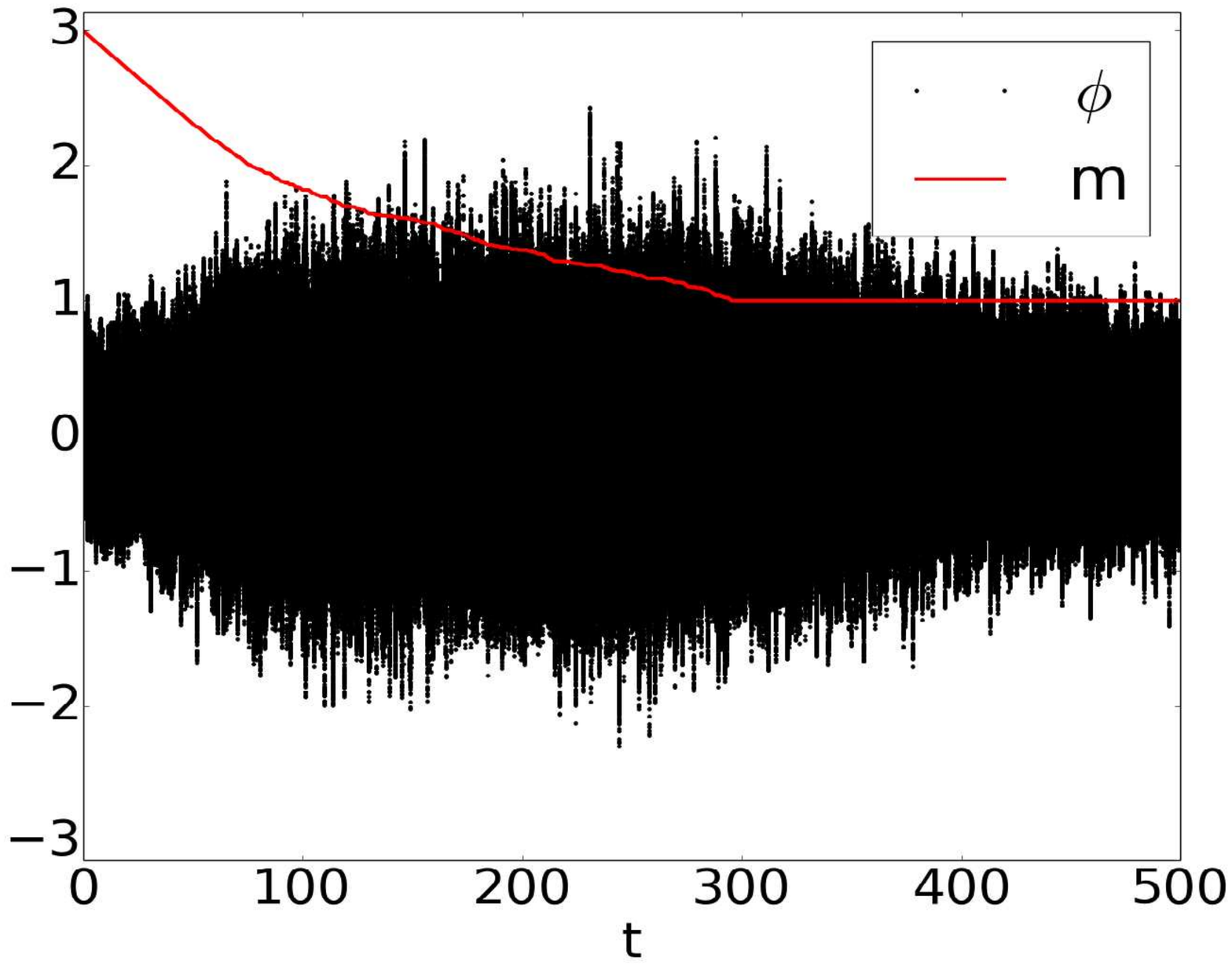}
\caption{\label{fig:moft} The argument of the phase of the fermion determinant 
along Langevin trajectory. With the Langevin time dependent quark mass 
(thin red curve) the fermion determinant does 
not circle the origin during the Langvin process. 
Here $\mu=1$, $N=20$, $N_f=2$, $\nu=0$ and the initial value of the quark 
mass $3$ is 
outside the cloud of Dirac eigenvalues. As the quark mass reaches the desired 
value $m=1$ it is kept fixed and the measurement of the chiral condensate 
can be performed.}
\end{figure}
\end{center}
\vspace{-35mm}

\subsection{Decreasing the quark mass with Langevin time}
\label{subsec:moft}

In large scale simulations it becomes exceedingly hard to invert the 
Dirac operator if the quark mass is inside the eigenvalue distribution. 
Moreover, in this case the fermion determinant is likely 
to circulate the origin frequently (when the quark mass is inside the 
Dirac eigenvalues the phase of the fermion determinant 
is distributed according to a Lorentzian distribution \cite{LSV}) and 
hence ignoring the cut of the logarithm is not 
necessarily justified \cite{MS}. In order to circumvent these issues    
it was proposed in \cite{DiracCL} to allow the quark mass to decrease 
with the Langevin time $t$. The proposal is to start from an initial 
value of the quark mass which is outside the Dirac 
eigenvalue distribution. With Langevin time the quark mass is then 
slowly decreased towards the desired value. In this way it is  
possible that the quark mass remains outside the distribution 
of the eigenvalues of the Dirac operator and that the fermion 
determinant does not circulate the origin at any given time throughout the 
Langevin process. Once the desired quark mass is reached all 
previous configurations are discarded from the measurement of the 
given observable.

Here we test this proposal within the Langevin process for chiral random 
matrix theory. We start the Langevin process on a random configuration from 
the original quenched ensemble (not complexified) and pick a value of 
the quark mass parameter which is safely outside the Dirac eigenvalues, 
i.e.~$m>2\mu^2$. The quark mass is reduced in steps proportional to the 
time step, unless the angle of the determinant has been above 1.5 within 
the last 1000 time steps. In that case we keep the quark mass constant 
to allow the Langevin dynamics to dampen the fluctuations of the phase 
of the determinant. This procedure is repeated until $m$ reaches the 
desired value, here 1, see figure \ref{fig:moft}.  
We observe that with Langevin time it is possible to decrease the 
quark mass such that the fermion determinant at no point during the 
Langevin trajectory circulates the origin. The potential problems with 
the log of the fermion determinant can therefore safely be ignored.

\section{Conclusions}
\label{sec:conc}

We have demonstrated that complex Langevin can simulate 
chiral random matrix theory at non-zero chemical potential even in the range corresponding to $\mu>m_\pi/2$.
The success of the complex Langevin method in chiral random matrix theory was 
established by {\sl 1)} a change of integration matrices and {\sl 2)} 
a polar parametrization of these variables before complexification. 
This choice of variables was 
inspired by taking the perspective of the eigenvalues of the Dirac 
operator evaluated on the complexified configurations. As shown in 
\cite{DiracCL} the Dirac spectrum of complex Langevin simulations 
must be vastly 
different form that with real configurations. In the application of 
complex Langevin to chiral random matrix theory, the natural solution is to
realize an effective anti-Hermitization of the 
Dirac operator through 
the complexification of the matrix elements. The choice of matrix integration 
variables and the polar parametrization of the elements hereof was handpicked 
to optimize 
the chance for complex Langevin to realize this effective anti-Hermitization.
Indeed, we have checked that the success of complex Langevin 
in chiral random matrix theory is established through an effective shift 
into the complex plane of the angular part of the matrix elements. This 
smoothly connects the initial non-Hermitian  
random matrix Dirac operator to an anti-Hermitian counterpart at large 
Langevin time. For a discussion of the possibility to realize a similar
scenario in the context of full QCD, see \cite{DiracCL}. 

As the chiral random matrix Dirac operator shares the chiral symmetries 
of the QCD Dirac operator it allows us to address several properties 
directly relevant for QCD. In particular, we have tested the proposal of 
\cite{DiracCL} in which the quark mass is initially Langevin time 
dependent: 
With adaptive step-size and a Langevin-time dependent quark mass we 
have demonstrated that it is possible to simulate the chiral random matrix 
theory at small mass (such that $m_\pi<2\mu$) without the determinant 
frequently circulating the origin. This minimizes the potential problems 
due to the non-holomorfic nature of the action in the presence of a fermion 
determinant. Furthermore, it ensures that inversions of the Dirac 
operator only appear with the quark mass outside 
the eigenvalues of the Dirac operator.

It would be most interesting to understand if the effect of a Langevin 
time dependent quark mass in complex Langevin simulations of full QCD is 
beneficial, in particular at low temperature and light quark mass. As an 
intermediate step a Langevin time dependent quark mass could also 
be implemented for the Thirring model \cite{Pawlowski}.  
Another possible direction is to use the eigenvalue representation 
of the chiral random matrix partition function as the basis for the 
Langevin process. Such an angle of approach has already lead to new 
insights for QCD in one dimension \cite{Aarts:2010gr}. 

\vspace{3mm}

\noi
{\bf Acknowledgements:} The work of AM is supported by the interdisciplinary 
UCPH 2016 grant 'Social Fabric' and the work of KS is supported by the Sapere 
Aude program of the Danish Council for Independent Research.

\end{document}